\documentstyle[prb,aps]{revtex}
\begin{document}
\draft
%@twocolumnfalse\endcsname
\title{{\bf Magnetic scaling in cuprate superconductors.}}
\author{\large V. Barzykin and D. Pines}
\address{ Department of Physics and
Science and Technology Center for Superconductivity, \\
University of Illinois at Urbana-Champaign, Urbana, Illinois 61801-3080 }
\maketitle
\begin{abstract}
We determine the magnetic phase diagram
for the YBa$_2$Cu$_3$O$_{6+x}$ and La$_{2-x}$Sr$_x$CuO$_4$
systems from various NMR experiments.
We discuss the possible interpretation of NMR and
neutron scattering experiments in these systems
in terms of both the non-linear $\sigma$-model of
nearly localized spins and a
nearly antiferromagnetic Fermi liquid description
of magnetically coupled quasiparticles.
We show for both the 2:1:4 and 1:2:3 systems that
bulk properties, such as the spin susceptibiltiy, and
probes at the antiferromagnetic wavevector $(\pi, \pi)$,
such as $^{63}T_1$, the $ ^{63}Cu$ spin relaxation time,
both display a crossover at a temperature $T_{cr}$, which
increases linearly with decreasing hole concentration,
from a non-universal regime to a $z=1$ scaling regime
characterized by spin pseudogap behavior. We pursue
the consequences of the ansatz that $T_{cr}$ corresponds to a fixed
value of the antiferromagnetic correlation length, $\xi$,
and show how this enables one to extract the magnitude and
temperature dependence of $\xi$ from measurements of $T_1$ alone.
We show that like $T_{cr}$, the temperature $T_*$ which marks
a crossover at low temperatures
from the $z=1$ scaling regime to a quantum disordered regime,
exhibits the same dependence on doping for the 2:1:4 and 1:2:3
systems, and so arrive at a unified description of magnetic behavior
in the cuprates, in which the determining factor is the planar
hole concentration. We apply our quantitative results for
YBa$_2$Cu$_3$O$_7$ to the recent neutron scattering experiments
of Fong {\em et al}, and show that the spin excitation near
$40 meV$ measured by them corresponds to a spin gap excitation,
which is overdamped in the normal state, but becomes visible in
the superconducting state.
\end{abstract}

\vspace{0.3in}
%]
%\narrowtext

\section{Introduction}
The exotic magnetic properties of high-temperature superconductors
have been extensively studied during the last few years, both because of
their intrinsic interest and in the hope that these might provide insight
in the physical origin of high-temperature superconductivity.
Although considerable progress has been made in understanding the
magnetic behavior of these materials, the basic issues of the doping
dependence of the magnetic phase diagram and the relationship between
transport properties and magnetic behavior in the normal state of the
superconducting cuprates have not yet been completely resolved.
In this paper we construct new magnetic phase diagrams for the
YBa$_2$Cu$_3$O$_{6+x}$ and La$_{2-x}$Sr$_x$CuO$_4$ system on the
basis of NMR experiments \cite{Slichter}. Our approach makes evident the
connection between the appearance of a spin pseudogap and magnetic
scaling, and enables us to deduce from NMR experiments of the $^{63}Cu$
spin-lattice relaxation rate the magnitude and temperature dependence
of the length, $\xi$, which characterizes the strength of the
antiferromagnetic correlations found in both underdoped and fully doped
cuprate superconductors.

It is well established that the undoped materials are described
by the 2D Heisenberg model with a fairly large exchange coupling $J \simeq
1550K$ \cite{Manousakis}.
The thermal fluctuations destroy long-range order at finite temperature
(the Hohenberg-Mermin-Wagner theorem), and the N\'eel transition observed in
undoped cuprates is solely due to the small, but finite interplanar coupling.
Chakravarty {\em et al} \cite{CHN} gave a strong indication that
the long-wavelength action of $s=1/2$ 2D
Heisenberg antiferromagnet is reduced to
the quantum non-linear sigma model, hereafter $\sigma$-model, characterized
by an action,
\begin{equation}
S_{eff} = \frac{\rho_S^{0}}{2}\,
\int_0^{\beta \hbar} d \tau \int d^2 x
\left[ ({\bf \nabla n})^2 +
\frac{1}{c_0^2}\, (\partial_{\tau}{\bf n})^2 \right].
\label{smodel}
\end{equation}
Here ${\bf n}$ is a three-component vector field, which describes the
local staggered magnetization, which is subject to the
constraint $| {\bf n} | = 1$, $\rho_S^{0}$ is the spin stiffness,
and $c_0$ is the spin wave velocity. There is a short-distance cutoff
$\Lambda^{-1}$ for the spatial integrals. The properties of
the $\sigma$-model are conveniently expressed in terms of the
coupling constant $g= \hbar c_0/\rho_S^0$.
Chakravarty {\em et al}  \cite{CHN}
identified three different regimes of behavior (Fig.\ref{phsigma}),
defined by the zero-temperature ordered-non-ordered phase transition
at $g_c=4 \pi/ \Lambda$.
According to the $\sigma$-model $g-T$ phase diagram,
for all coupling constants there exists a
high-temperature Quantum Critical (QC) regime
($(\rho_S, \Delta) \lesssim T \lesssim J$) in which the characteristic
energy scale is set by temperature.
At lower temperatures, depending on the coupling constant, $g$,
the correlation length either grows exponentially,
as in the Renormalized Classical
(RC) regime (for $g < g_c$) or stays constant, as in
the Quantum Disordered (QD) regime
(for $g > g_c$).
For the states without long-range order
($g>g_c$ or $T>0$)
the Goldstone mode is
eliminated, which leads to a finite gap in the spin wave spectrum
$\Delta = c/\xi$, where $c$ is the spin wave velocity,
and $\xi$ the correlation length.

Doping changes the ground state of the cuprates from
a Ne\'el-ordered state to a highly-correlated ground state
without long-range order \cite{Anderson}. It was first argued by
Sachdev and Ye \cite{Sachdev:Ye} that the behavior of the cuprates for
small doping is controlled by the zero-temperature critical point of
the $\sigma$-model leading to the dynamic critical exponent $z=1$.
This critical point does not
necessarily correspond to the metal-insulator transition.
The Quantum Critical (QC) $z=1$ transition exhibited by the $\sigma$-model
was then analysed in detail by Chubukov {\em et al}
\cite{Chubukov:Sachdev}, who emphasized its potential importance
for the magnetic phase diagram of the
cuprate superconductors.

On the basis of NMR experiments of Imai {\em et al}
\cite{Imai:Slichter} and Takigawa \cite{Takigawa},
Sokol and Pines \cite{Sokol:Pines} then
argued that the $\sigma$-model, QC $z=1$, description of the spin
fluctuations continues to be valid at the comparatively large doping
levels which are appropriate to the underdoped cuprate
superconductors.
They suggested that a spin wave description is valid
at high temperatures for these systems, while
at higher doping levels, such as are found in YBa$_2$Cu$_3$O$_7$,
the particle-hole excitations act to completely damp the
spin wave spectrum, so that the normal state of YBa$_2$Cu$_3$O$_7$
is best described using the  nearly antiferromagnetic Fermi liquid
(hereafter NAFL) model which had been successfully applied to the
analysis of NMR experiments in both its normal and superconducting
state \cite{MMP,Moriya,TPL}. In this mean field model the
spin fluctuations at $(\pi, \pi)$ are characterized by a
relaxational mode which varies as $\xi^{-2}$; thus it resembles a system
with the dynamical critical exponent, $z=2$.
In Ref. \cite{BPST} the mean-field formula for the crossover
from $z=1$ to $z=2$  was obtained
by incorporating the quasiparticle damping in the $\sigma$-model
expression for the spin susceptibility, and  different experiments
were analysed using  this formula.
A major step towards the description
of the $z=1$ to $z=2$ crossover  has been taken recently
by Sachdev {\em et al} \cite{SCS}, who explicitly derive
the crossover scaling functions of the spin-fermion model using
a $1/N$ expansion.

A very general analysis of both NAFL and nearly ferromagnetic
Fermi liquid behavior has been carried out by Millis and Ioffe
\cite{Millis,Millis:Ioffe}
while Monthoux and Pines \cite{Monthoux:Pines}, hereafter MP,
have discussed the
possibility that a NAFL approach, in which non-linear feedback
from the magnetic interaction between quasiparticles
acts to bring about a spin pseudogap in the quasiparticle spectrum,
might yield the functional equivalent of QC $z=1$ behavior over a
broad temperature region in the underdoped cuprates, with overdamped
spin waves arising because the magnetic energy, $\omega_J \sim$
a few $J$, and the quasiparticle Fermi energy are on approximately
the same scale. They suggested that the onset of the pseudogap would
occur at a certain value of the correlation length, which is
independent of doping.

In the present paper we extend previous analyses of NMR experiments
\cite{Sokol:Pines,BPST} on the metallic low temperature normal state
regimes of La$_{2-x}$Sr$_x$CuO$_4$ and YBa$_2$Cu$_3$O$_{6+x}$
systems and present revised magnetic phase diagrams (Fig. \ref{phase}
and Fig. \ref{phase:La}) which make explicit the presence of both
an upper limit, $T_{cr}$, and a lower limit, $T_*$, to QC $z=1$
scaling behavior in these systems. We show that $T_{cr}$ both marks
the onset of spin pseudogap behavior,
a reduction of the effective quasiparticle
density of states, found first in bulk susceptibility \cite{Johnston}
and Knight shift \cite{Alloul} experiments, and
subsequently in specific heat experiments \cite{Loram},
and the crossover from a high-temperature
non-universal regime to a low-temperature universal scaling regime.
This scaling regime has dynamical critical exponent $z=1$.
We argue that this crossover happens at a fixed value of the
correlation length, $\xi_{cr} \simeq 2$, and that $T_{cr}$ is
close to $T_{\chi}$, the temperature which marks the maximum
of the spin susceptibility. We use the results of
NMR experiments to determine the region of applicability
of scaling and find that $T_{cr}$ varies nearly linearly
with doping level over a region which extends from the AF
insulators, YBa$_2$Cu$_3$O$_6$ and La$_2$CuO$_4$, to YBa$_2$Cu$_3$O$_7$
and the overdoped system La$_{1.76}$Sr$_{0.24}$CuO$_4$.
We find (see Fig. \ref{spg}) that when the crossover temperature,
$T_{cr}$, is plotted as a function of planar hole doping, within
experimental uncertainties, one finds the same dependence on
doping for the 1:2:3 and 2:1:4 families. This demonstrates that the
onset of magnetic scaling is determined only by hole concentration
(band-structure effects are unimportant) and that bilayer coupling
plays little or no role in determining spin pseudogap and scaling
behavior.

The QC-QD crossover at $T_*$, which
is predicted by the $\sigma$-model
at lower temperatures, is also identified from the NMR data.
Our identification of the experimentally determined
``spin pseudogap'' temperature, $T_{cr}$, with a fixed value of
$\xi$ then allows us to use NMR measurements of the $^{63}Cu$
spin-lattice relaxation rate to determine the correlation length
and other parameters which characterize low frequency magnetic
behavior as a function of temperature
for both the YBa$_2$Cu$_3$O$_{6+x}$  and La$_{2-x}$Sr$_x$CuO$_4$ families.
We establish thereby the details of the variation with doping of the
correlation length.

There exists as well
an intimate relationship between spin and charge response of the
underdoped cuprate superconductors. Ito {\em et al} \cite{Ito}
showed for YBa$_2$Cu$_3$O$_{6.63}$ that below a temperature,
$T_{\rho}$, the planar resistivity
ceases to exhibit a linear in T behavior, while the Hall effect
likewise changes character at this temperature; a similar conclusion
for YBa$_2$Cu$_4$O$_8$ was reached by Bucher {\em et al} \cite{Bucher}.
We find that within experimental error, $T_{\rho} = T_*$.
In the La$_{2-x}$Sr$_x$CuO$_4$ system
Hwang {\em et al} \cite{Battlogg} have shown that
for $x \ge 0.15$, the Hall coefficient,
$R_H(T)$, follows scaling behavior, with a characteristic temperature,
$T_H$, which not only exhibits the same strong dependence on doping
level that is found for $T_{cr}$, but which possesses very nearly
the same magnitude as the $T_{\chi}$ determined from the maximum in the
spin susceptibility. For this same system, with $x \le 0.14$,
Nakano {\em et al} \cite{Nakano} find that below $T_{\chi}$,
the resistivity ceases to be linear in T.
Thus both the lower crossover temperature, $T_*$,
and the upper crossover temperature, $T_{cr}$, which mark the limits
of the QC $z=1$ scaling behavior we identify in low frequency
magnetic measurements, possess counterparts in transport
measurements.

Although there is indirect experimental evidence for their
existence from NMR experiments,
propagating spin waves
for the underdoped cuprates have not been observed
in neutron scattering experiments on the normal state.
We show, on the basis of the parameters we infer from
NMR experiments, that just such a result is to be expected -
that spin waves are substantially overdamped by quasiparticle interactions
in the normal state. Since this damping is markedly reduced in the
superconducting state, we conclude that spin
wave excitations should become visible in the superconducting state,
and we present arguments which suggest that this provides a quantitative
explanation for the recent experiments of Fong {\em et al} \cite{Fong}
on inelastic neutron scattering from YBa$_2$Cu$_3$O$_7$.

Our paper is organized as follows.
In Section II we review the one-component approach to
the interpretation of NMR measurements.
In Section III we use this to
analyse the
current experimental situation for the magnetic measurements in the
underdoped YBa$_2$Cu$_3$O$_{6+x}$ and La$_{2-x}$Sr$_x$CuO$_4$ families, and
on the basis of this analysis we arrive at our new
magnetic phase diagram for these systems.
In Section IV we discuss possible theoretical models, and
the form for the spin susceptibility at high energy transfer.
We discuss the application of the scaling analysis
to inelastic neutron scattering experiments in section V, and
in Section VI we present our conclusions.

\section{The one-component approach to NMR experiments.}

We start by giving a brief overview of the NMR
experiments and their interpretation in one-component
nearly antiferromagnetic Fermi liquid (NAFL) theory.
The interested reader will find a more detailed discussion in the
excellent recent review of
NMR measurements in the cuprate superconductors by
Slichter \cite{Slichter}.

Nuclear spins probe the local environment. The Knight shift provides a
measure of the uniform magnetic susceptibility at a particular nuclear
site, while the measurements of the spin-lattice relaxation rates yield
information on the imaginary and real parts of the dynamic
spin susceptibility $\chi({\bf q}, \omega)$.  The single-component
description of the NMR experiments in YBa$_2$Cu$_3$O$_7$ has its basic
justification in the observation by Alloul {\em et al} \cite{Alloul} and
Takigawa {\em et al} \cite{Takig} that the $^{63}Cu$,
$^{17}O$ and $^{89}Y$ Knight shifts  see
the same spin susceptibility. Since the $^{63}Cu$
spin-lattice relaxation time both has anomalous temperature dependence
and is much shorter than that for the $^{17}O$ and $^{89}Y$ nuclei,
it was proposed\cite{Mila:Rice:Shastry} that the single magnetic
component formed by the
system of planar Cu$^{2+}$ spins and holes mainly resides on the
planar copper sites. The Shastry-Mila-Rice (hereafter SMR) Hamiltonian
\cite{Mila:Rice:Shastry} which describes
the coupling of the Cu$^{2+}$ spins to the various nuclei in
YBa$_2$Cu$_3$O$_7$, has the  form:
\begin{eqnarray}
H_{MRS} & = & \ ^{63}{I}_{\alpha}({\bf r}_i) \left[\sum_{\beta}
A^{\alpha \beta} S_{\beta}({\bf r}_i) + B
\sum_j^{nn} S_{\alpha} ({\bf r}_j) \right] + \nonumber \\
& & + \ ^{17} I_{\alpha}({\bf r}_i) C^{\alpha \beta}
\sum_{j \beta}^{nn} S_{\beta}({\bf r}_j) +
\nonumber \\
 & & + \ ^{89} I_{\alpha}({\bf r}_i) D \sum_j^{nn} S_{\alpha}({\bf r}_j),
\label{MR}
\end{eqnarray}
where $A_{\alpha \beta}$ is the tensor for the direct, on-site
coupling of the $^{63}Cu$ nuclei to the $Cu^{2+}$ spins,
B is the strength of the transferred hyperfine coupling of the $^{63}Cu$
nuclear spin to the four
nearest neighbor copper spins, while $^{17}O$ and $^{89}Y$ nuclei
see only their nearest neighbor $Cu^{2+}$ spins through the
transferred hyperfine interactions $C^{\alpha \beta}$ and D.
It follows from this Hamiltonian that different nuclei probe different
regions in momentum space of $\chi({\bf q}, \omega \rightarrow 0)$.
Using eq.(\ref{MR}), it is straightforward to express the spin contribution
to the Knight shifts for the various nuclei\cite{MMP}:
\begin{eqnarray}
\label{Kn}
^{63}K^S_{\parallel} & = &
\frac{(A_{\parallel} + 4 B) \chi_0}{^{63}\gamma_n \gamma_e \hbar^2} \\
^{63}K^S_{\perp} & = &
\frac{(A_{\perp} + 4 B) \chi_0}{^{63}\gamma_n \gamma_e \hbar^2} \nonumber \\
^{17}K^S_{\beta} & = &
\frac{2 C_{\beta} \chi_0}{^{17}\gamma_n \gamma_e \hbar^2} \nonumber \\
^{89}K^S  & = &
\frac{8 D \chi_0}{^{89}\gamma_n \gamma_e \hbar^2}. \nonumber
\end{eqnarray}
Here the $\gamma_n$ are various nuclei gyromagnetic ratios,
$\gamma_e$ is the electron gyromagnetic ratio, $\chi_0$ is the
temperature-dependent static spin susceptibility, while the indices $\parallel$
and $\perp$ for copper nuclei refer to the direction of the applied static
magnetic field with respect to the axis perpendicular to the $Cu-O$ planes.
The index $\beta$ for the oxygen nuclei can be either $\parallel$ or $\perp$.
The spin-lattice relaxation time, $(^{\alpha}T_1)_{\beta}$,
for a nucleus $\alpha$ responding to a magnetic field in the
$\beta$ direction, is:
\begin{equation}
\label{T1}
^{\alpha}T_{1 \beta}^{-1}(T) = \frac{k_B T}{2 \mu_B^2 \hbar^2 \omega}
\sum_q \ ^{\alpha}F_{\beta}({\bf q}) \chi''({\bf q}, \omega \rightarrow 0),
\end{equation}
where the form factors $^{\alpha}F_{\beta}({\bf q})$ are given by:
\begin{eqnarray}
\label{formfactors}
^{63}F_{\parallel} & = &
\left[A_{\perp} + 2 B (cos(q_x a) + cos(q_y a)) \right]^2 \\
^{63}F_{\perp} & = & \frac{1}{2}\,
\left[\ ^{63}F_{\parallel} + \ ^{63}F_{eff} \right]  \nonumber \\
^{63}F_{eff} & = & \left[A_{\parallel} +
2 B (cos(q_x a) + cos(q_y a)) \right]^2 \nonumber \\
^{17}F_{\beta} & = & 2 C_{\beta}^2
\left[1 + \frac{1}{2}\,(cos(q_x a) + cos(q_y a)) \right] \nonumber \\
^{89}F_{ISO} & = & 16 D^2 cos^2(q_z a/2) [1+cos(q_x a)][1+cos(q_y a)].
\nonumber
\end{eqnarray}
The form factor $^{63}F_{eff}$ is the filter for the $^{63}Cu$ spin-echo
decay time, $^{63}T_{2G}$\cite{Thelen:Pines}:
\begin{eqnarray}
^{63}T_{2 G}^{-2}(T) & = & \left(\frac{0.69}{128}\right)^{1/2}
(^{63}\gamma_n)^2 \left\{\frac{1}{N}\, \sum_{\bf q}
F_{eff}({\bf q})^2 [\chi'({\bf q}]^2 - \right. \nonumber \\
 & & - \left.
\left[\frac{1}{N}\sum_{\bf q} F_{eff}({\bf q}) \chi'({\bf q}) \right]^2
\right\}
\label{T2G}
\end{eqnarray}
The hyperfine coupling constants $A_{\perp}$, $A_{\parallel}$,
$C$, and $D$, which enter the SMR Hamiltonian,
Eq. (\ref{MR}), were determined using the Knight shift measurements
in YBa$_2$Cu$_3$O$_7$ and the frequency of the antiferromagnetic
resonance in YBa$_2$Cu$_3$O$_6$\cite{MMP,MPT}. The momentum dependence of the
form factors is shown in Fig. \ref{fofa}. A striking difference
of the temperature dependence and the magnitude of the spin-lattice
relaxation on the $^{63}Cu$, $^{17}O$, and $^{89}Y$ nuclei led
Millis {\em et al} \cite{MMP} to the conclusion that the magnetic
response function is strongly peaked at the antiferromagnetic
wave vector, as might be expected if one were close to
an antiferromagnetic singularity.
Since the form factors for $^{17}O$ and $^{89}Y$ nuclei vanish at
${\bf Q}=(\pi/a, \pi/a)$, the difference in the magnitude and temperature
dependence of $^{63}Cu$ and other relaxation rates can be
easily explained. Once
the assumption of strong enhancement of the magnetic response function
at the antiferromagnetic wave vector is made, the  $^{63}Cu$  spin-lattice
relaxation anisotropy ratio provides a self-consistency check on the
values of the hyperfine constants,
\begin{equation}
^{63}R = \frac{T_{1 \parallel}}{T_{1 \perp}} \, =
\frac{^{63}F_{\perp}({\bf Q})}{^{63}F_{\parallel}({\bf Q})} \simeq 3.7
\end{equation}
The values of the hyperfine constants,
\begin{eqnarray}
\label{const}
B &=& 3.82 \times 10^{-7} eV, \ \ A_{\parallel} = -4 B, \\ \nonumber
A_{\perp} &=& 0.84 B, \ \ C_{\parallel} = 0.91 B,
\end{eqnarray}
satisfy this check. Moreover, these hyperfine coupling constants remain
unchanged (at the 5\verb+%+ to 10\verb+%+ level)  with the variation
of the doping level or for
different cuprate materials, as might be expected from
quantum chemical arguments.  Thus these constants can be regarded as
the same for both the  YBa$_2$Cu$_3$O$_{6+x}$ and La$_{2-x}$Sr$_x$CuO$_4$
families.

We follow Millis {\em et al} \cite{MMP} and
Monien {\em et al} \cite{MPT}
and use a response function consisting of
an anomalous part,  $\chi_{MMP}({\bf q}, \omega)$, and a Fermi-liquid part,
$\chi_{FL}({\bf q}, \omega)$,
\begin{equation}
\chi({\bf q}, \omega) = \chi_{MMP}({\bf q}, \omega) +
\chi_{FL}({\bf q}, \omega).
\label{MPTeq}
\end{equation}
The expression used by
Millis {\em et al} \cite{MMP} for the anomalous part can be written
in the mean-field form:
\begin{equation}
\chi({\bf q}, \omega)_{MMP} = \frac{\chi_Q}{1 + ({\bf q} - {\bf Q})^2 \xi^2 -
i \omega/\omega_{SF}},
\label{MMP}
\end{equation}
where
\begin{equation}
\chi_Q = \alpha \xi^2
\label{Sfac}
\end{equation}
Here $\omega_{SF}$ is the characteristic frequency of the
spin fluctuations, $\xi$ is the correlation length, $\alpha$ is
a scale factor.
Due to the strong antiferromagnetic enhancement, the anomalous contribution
to the copper relaxation rate dominates at all doping levels.
However, since the form factors
for other nuclei are zero at the antiferromagnetic wave vector, the Fermi
liquid (Korringa-like) contribution to their spin-lattice relaxation
rate becomes important.

In the limit of long
correlation lengths the expressions for the $^{63}Cu$ relaxation rates
eqs (\ref{T1}), (\ref{T2G}) are considerably simplified \cite{Thelen:Pines}:
\begin{equation}
^{63}(T_1 T) = 132 \ (s \cdot K)/eV^2 \ \omega_{SF}/\alpha,
\label{T1x}
\end{equation}
while
\begin{equation}
^{63}\frac{1}{T_{2G}}\, = 295 \ (eV/s) \ \alpha \xi
\label{T2x}
\end{equation}
Thus, up to a scale factor $\alpha$,
$^{63}(T_1 T)$ provides a direct measurement of $\omega_{SF}$,
while $^{63}T_{2G}$ provides a measurement of the correlation length.
Moreover, the ratios:
\begin{eqnarray}
\frac{^{63}T_1 T}{^{63}T_{2G}} & =  & 3.9 \times 10^4 (K/eV)
(\omega_{SF} \xi = c')  \\      \nonumber
\frac{^{63}T_1 T}{^{63}T_{2G}^2} & = & 1.15 \times 10^7 (K/s)
(\alpha \omega_{SF} \xi^2 = \chi_Q \omega_{SF})  \\
\label{sc}
\end{eqnarray}
tell us about the scaling laws \cite{Sokol:Pines}, if any, obeyed
by the low frequency magnetic excitations.
Thus, as emphasized in Refs \cite{Sokol:Pines}, \cite{Millis:Monien},
if $^{63}T_1 T/ T_{2G}$ is independent
of temperature, one has QC $z=1$ scaling behavior, and the doping-dependent
ratio, $\omega_{SF} \xi = c'$, is determined. On the other
hand, when $^{63}T_1 T/ ^{63}T_{2G}^2 = const$, $\omega_{SF}$ displays
either non-universal mean field or QC $z=2$ scaling behavior, and the
product $\chi_Q \omega_{SF}$ is independent of temperature.

\section{Experimental phase diagrams.}

In this section we use the  NMR results for $^{63}T_1$, $^{63}T_{2G}$
and the Knight shift on the underdoped 1-2-3 and the 2-1-4 families
to establish the complete experimental
magnetic phase diagram for these systems.
We first review briefly the properties of the regimes which might
be relevant to the experimental situation in the
metallic cuprates \cite{Sokol:Pines}:

\bigskip
\noindent
{\Large$\bullet$} The QC, $z=1$ regime. In this regime
the spin-spin correlator takes the form:
\begin{equation}
\chi({\bf q}, \omega) = \xi^{2-\eta} F({\bf q} \xi, \omega \xi),
\label{QCSc}
\end{equation}
and the inverse correlation length varies linearly with temperature:
$1/\xi= a T + b$, where in the $\sigma$-model \cite{Chubukov:Sachdev}
$a$ and $b$ are computable universal constants.
The key signature of the QC $z=1$ regime in NMR measurements
\cite{Sokol:Pines} is that
the ratio $^{63}T_1T/{^{63}T_{2G}} \propto \omega_{SF} \xi = c'$
is temperature-independent (Eq.(\ref{sc})), and $c'$ is,
for the $\sigma$-model,
proportional to the spin-wave
velocity $c$, $c' \simeq 0.55 c$.
In the QC regime the characteristic frequency $\omega_{SF}$ is
linear in temperature. The bulk magnetic susceptibility should also
vary linearly with temperature,
\begin{equation}
\chi(T) = \chi_{univ}(T) + \chi_0,
\label{chiun}
\end{equation}
where $\chi_{univ}(T)$ is a universal linear contribution,
computable in the $\sigma$-model\cite{Chubukov:Sachdev}:
\begin{equation}
\chi_{univ}(T) = \left(\frac{g \mu_B}{\hbar c}\right)^2 (0.34 k_B T - 0.137
\Delta(T=0)),
\label{chiuni}
\end{equation}
while $\chi_0$, the fermionic contribution,
is temperature-independent.

\bigskip
\noindent
{\Large$\bullet$} The Quantum Critical, $z=2$, regime\cite{MMP,Millis}.
The characteristic energy scale in this regime is again set by temperature,
so that at high enough temperatures $\omega_{SF} \propto T$. However,
since $z=2$, $\omega_{SF} \propto \xi^{-2}$. As a result,
in this regime ${^{63}T_1T}/{^{63}T_{2G}^2} = const$, and
at high temperatures the inverse squared correlation length should
obey the Curie-Weiss law, $\xi^{-2} = a T + b$.

\bigskip
\noindent
{\Large$\bullet$}
The QD regime. In this regime the correlation length (and
$^{63}T_{2G}$) saturates. If $z=2$, the damping frequency $\omega_{SF}$
(and $^{63}T_1T$) should  saturate at zero temperature to a certain value,
determined by the fermionic damping \cite{BPST,SCS}. In case of the $z=1$
$\sigma$-model, $\omega_{SF}$ in the QD regime
increases exponentially with decreasing
temperature\cite{Chubukov:Sachdev}.

\bigskip
\noindent
{\Large$\bullet$} Non-scaling regime. In this regime lattice effects
are appreciable, and a scaling analysis is inapplicable. As a result,
the calculations in this regime are model-dependent. A subclass of the
non-scaling regime is the mean field NAFL at high doping, where
Eq.(\ref{MMP}) can be used.

In order to describe the $z=2$ regime, the $\sigma$-model
Eq.(\ref{smodel}) should be generalized to include
fermionic damping \cite{SCS}. The effect of this damping is that at
low enough temperatures one always should find a crossover
from the $z=1$ to $z=2$ scaling regime. This crossover happens either
in the QC or the QD regime, depending on the value of the fermionic damping.
We briefly discuss this model in Section IV.

We begin our analysis of magnetic behavior by considering
La$_{2-x}$Sr$_x$CuO$_4$, where
the spin susceptibility $\chi_0$ has been measured both
in the bulk experiments\cite{Johnston,Nakano}
and in NMR Knight shift probes\cite{Alloul,Takig}, and $^{63}T_1$
measurements have been carried out for many different
doping levels\cite{Imai:Slichter,Kitaoka}.
The decrease in the spin susceptibility as the temperature is lowered over
a broad temperature region was the first identification  in the experimental
literature
of the  ``spin pseudogap'' phenomenon.
At a temperature $T_{\chi}$, the spin susceptibility (or the
$^{63}Cu$ Knight shift) which at high
temperatures typically decreases somewhat with increasing temperature,
as shown in Figs \ref{chi:La} and \ref{knight}, changes behavior; below
$T_{\chi}$ it
decreases as temperature is lowered.  Loram {\em et al}
\cite{Loram} have carried out specific heat measurements which
show that this decrease in $\chi_0$ is matched by a corresponding
decrease in the quasiparticle density of states, $N_0(T)$.
Johnston \cite{Johnston} and subsequently Nakano {\em et al} \cite{Nakano}
also found that the bulk magnetic susceptibility could be expressed in a
scaling form:
\begin{equation}
\chi(T) = \chi_0 + \kappa \chi(T/T_{max})
\label{exsc}
\end{equation}
Nakano {\em et al} \cite{Nakano} used $\chi_0$, $\kappa$ and
$T_{max}$ as fitting parameters. They found that the
temperature-independent term $\chi_0$ is doping-dependent; it
therefore cannot be ascribed to the Van-Vleck contribution.
This scaling behavior is easily explained
if one identifies $\chi_0$ as the ``fermionic'' part, and
$\kappa \chi(T/T_{max})$ as the ``spin'' part. Eq.(\ref{exsc})
is then quite similar to Eq.(\ref{chiun}), the result expected
from the scaling theory.
However, it should be emphasized that from our point of view, because of
large lattice corrections, scaling ends above some temperature, $T_{cr}$.
 To preserve Eq.(\ref{exsc}), the lattice corrections should
enter ``universally'', i. e. preserving the relationship
Eq.(\ref{chiun}), while breaking the $\sigma$-model linear
temperature dependence of $\chi_{univ}(T)$.

Quite generally, for a $z=1$ scaling regime which extends
between a lower crossover temperature, $T_*$, and an upper
crossover temperature, $T_{cr}$, one expects scaling behavior
to manifest itself at both long wavelengths (i. e. in $\chi_0(T)$ )
and at the wavevectors around ${\bf Q}$ which determine both
$^{63}T_1$ and  $^{63}T_{2G}$, with both
$\chi_0(T)$ and $^{63}T_1T$ exhibiting a linear temperature dependence.
$T_{cr}$ marks the onset of scaling behavior, while $T_*$ marks beginning
of the crossover from the QC ($z=1$) to the
QD regime, predicted in the scaling theory\cite{CHN}, which
has been previously identified in the NMR
$^{63}T_1$ and $^{63}T_{2G}$ data \cite{Sokol:Pines,BPST}.
This is the temperature at which $^{63}T_1 T$ and $^{63}T_{2G}$
stop being linear in $T$, as is expected for the QC $z=1$
regime. We propose that $T_{cr}$ marks the crossover from scaling
and spin pseudo-gap behavior to non-universal behavior.
Inspection of Fig.\ref{chi:La} then show that $T_{cr}$ cannot be
far from $T_{\chi}$; in other words, the maximum in $\chi_0(T)$
is reached at very nearly the same temperature at which scaling
begins. Inspection of Fig.\ref{QD:La}  shows that the crossover at $T_{cr}$ is
also visible in $^{63}T_1T$, as a change in slope; for the various samples
for which both $\chi_0(T)$ and $^{63}T_1$ measurements have been carried
out, the two independent methods (one long wavelength, one short wave length)
of determining $T_{cr}$ are seen to be in good agreement with one another.
The lower crossover, at $T_*$, is readily visible in the
$^{63}T_1$ data of Ohsugi {\em et al} \cite{Kitaoka} and Imai {\em et al}
\cite{Imai:Slichter} shown in Fig. \ref{QD:La}. It is
also clearly visible in $\chi_0(T)$ measurements for $Sr$ doping levels,
$0.15$ and below, as marking the onset of a more rapid fall-off in
$\chi_0(T)$ as the temperature is further decreased.
Our experimentally determined values of $T_*$ and $T_{cr}$ are given
in Fig. \ref{phase:La}; $T_*$ is seen to vary comparatively slowly with
hole concentration; it possesses a maximum for doping levels in the vicinity
of La$_{1.85}$Sr$_{0.15}$CuO$_4$. $T_{cr}$ on the other hand varies rapidly
with hole concentration, for doping levels less than
La$_{1.8}$Sr$_{0.2}$CuO$_4$.
We indicate by a solid line our linear extrapolation of $T_{cr}$ to doping
levels less than La$_{1.85}$Sr$_{0.15}$CuO$_4$ (where high temperature
measurements
of $^{63}T_1$ have not yet been carried out); we note that this extrapolation
suggests a crossover temperature of $\sim 1200 K$ for the AF insulator,
La$_2$CuO$_4$, not far from that found from the high-temperature series studies
of the 2D Heisenberg antiferromagnet\cite{Norbert}.

Spin pseudogap and scaling behavior has also been found in transport
measurements on the 2-1-4 system.
Hwang {\em et al} \cite{Battlogg} and Nakano {\em et al}
\cite{Nakano} analysed
the data for the bulk magnetic susceptibility,
resistivity \cite{Nakano}, and Hall effect\cite{Nakano,Battlogg}
measurements
for this system.
They identified $T_{\chi}$ from the
spin susceptibility, and found similar characteristic
temperatures, $T_{\rho}$ and $T_H$, from an analysis of change with temperature
of the resistivity and Hall effect.
It is not necessary, from our point of view, that the
temperatures identified  using these criteria be exactly the same. However,
as may be seen in Fig. \ref{spg},
where our suggested doping dependence of
$T_{cr}$ is compared with $T_{\chi}$, $T_{\rho}$, and T$_{R_H}$,
these respective temperatures are very close to each other; a
closeness which makes evident the
inseparability of spin and charge behavior in this system.
We note that due to the vague nature of the
crossovers large errors of $\sim 15 \verb+%+$ are involved in our
estimates for the values of crossover temperatures. Errors may also occur
because the doping is not always exactly known,
while different measurements are
often performed on different samples.

Having established that the phenomena of scaling and spin pseudogap behavior
are intimately related, and that both begin at $T_{cr}$, we next explore the
possibility suggested by Monthoux and Pines \cite{Monthoux:Pines}, that
$T_{cr}$
is associated with a critical value of the correlation length, $\xi_{cr}$,
which is independent of doping.  One reaches the same conclusion on the basis
of
the $\sigma$-model, where one argues that lattice corrections bring about an
end
to scaling for $\xi > \sim 1$. We make the ansatz that $\xi_{cr} \simeq 2$.
Our reason for choosing this value is that one finds that in both the
insulating and nearly fully-doped samples, the onset of spin pseudogap behavior
coincides with $\xi \simeq 2$. Thus in the insulator, the bulk susceptibility
reaches its maximum at $T \sim 1400 K$ \cite{Johnston}, not far from the
temperature ($T_{cr} \simeq 1000K$) at which $\xi = 2$ according to high
temperature series studies\cite{Norbert}, while, as we shall see below,
at the opposite end of the doping scale, $\xi \simeq 2$ at the temperature
(125 K) which marks the onset of spin pseudogap behavior in almost
fully-doped YBa$_2$Cu$_3$O$_7$.

Unfortunately, $^{63}T_{2G}$ data is not yet available in the
metallic 2:1:4 samples, so that one cannot verify directly (as we shall be
able to do for their 1:2:3 counterparts) that the scaling law,
$\omega_{SF} \xi = c'$, a constant, is obeyed  for these systems. However,
on the assumption that it is valid , and moreover that $\xi_{cr} =2$, we can
determine the correlation length in the QC regime based on the NMR $^{63}T_1$
measurements only. To see this, note that between $T_*$ and $T_{cr}$ we have:
\begin{equation}
^{63}T_1(T) T = 132 (s \cdot K) \frac{c'}{\alpha \xi(T)} \ \ (T_* \le T \le
T_{cr}),
\label{TTc}
\end{equation}
so that a knowledge of $T_{cr}$ from another experiment  (e. g. the maximum in
$\chi_0(T)$) serves to fix $c'/\alpha$, and hence $\xi(T)$.
The doping dependence of
$c'/\alpha$ can also be determined from $^{63}T_1T$ only, since
\begin{equation}
\frac{c'}{\alpha}\, = ^{63}(T_1(T_{cr}) T_{cr}) \frac{eV^2}{66 s \cdot K}
\label{calp}
\end{equation}
The correlation length as a function of temperature and $c'/\alpha$
for different doping levels are shown in Figs \ref{cor:La},\ref{ca:La}.

It is instructive to consider the extent to which the values of $\xi(T)$
we have determined from $^{63}T_1$ display the universal behavior predicted
by the $\sigma$-model for the $z=1$ scaling regime, $\xi^{-1} = a + b T$,
for the case in which $c$ and $c'$ do not vary with doping.
We see from Table \ref{dat}, the coefficients $a$ and $b$ are not
universal, which is a clear indication that $c$, $c'$, and hence $\alpha$,
must be doping-dependent in this system. On the other hand,
we find that the slope of $^{63}T_1T$ is nearly doping-independent
between $T_*$ and $T_{cr}$
($^{63}T_1T (ms \cdot K) =  28 +0.027 T$ for La$_{1.9}$Sr$_{0.1}$CuO$_4$
and $^{63}T_1T (ms \cdot K) =  10 +0.025 T$  for
La$_{1.76}$Sr$_{0.24}$CuO$_4$),
which is what scaling theory for a doping-dependent $c$ would predict.
We note that our assumption that $\xi \simeq 2$ at $T_{cr}$ means that
$\xi^{-1}$
may be written in a simple form,
\begin{equation}
\frac{1}{\xi}\, = a + (\frac{1}{2}\, - a )(T/T_{cr}),
\label{xi-tcr}
\end{equation}
which experiments given in Table \ref{dat} satisfy.

We turn next to the YBa$_2$Cu$_3$O$_{6+x}$ family. A complete
analysis can be done for the compounds  YBa$_2$Cu$_3$O$_{6.63}$,
YBa$_2$Cu$_4$O$_8$ and YBa$_2$Cu$_3$O$_7$, where $^{63}T_1$,
$^{63}T_{2G}$, and Knight shift data are available.
A less complete analysis, comparable to that given above for the
2-1-4 system, can be given for the Pr-doped
YBa$_2$Cu$_3$O$_7$, where only $^{63}T_1$ and Knight shift experiments
have been carried out.
We start by analysing the Knight shift experiments in
the $Pr$-doped
compounds, Y$_{1-x}$Pr$_x$Ba$_2$Cu$_3$O$_7$ \cite{Reyes},
with x=0.05, 0.1, and 0.15.
We assume that $Pr$ substitution acts to remove, on average,
$1/2$ hole per plane for each substituted $Pr$ atom.
Thus the planar hole concentration in Y$_{1.95}$Pr$_{0.05}$Cu$_3$O$_7$
is assumed to be $0.025$ less than that of the ``host'' material,
YBa$_2$Cu$_3$O$_7$. But, as noted above, the hole concentration
of that material is not precisely known, so that a corresponding
degree of uncertainty affects the doping dependence we propose
for this system.
The spin part of the
bulk magnetic susceptibility we obtain for these
compounds (in states/eV) is shown in Fig.\ref{knight}.
As we have done for the 2-1-4 system,
we identify the
onset of scaling behavior, $T_{cr}$, with the maximum, $T_{\chi}$,
of the curve $\chi_0(T)$. We note that the bulk spin
susceptibility varies linearly with temperature between $T_*$
and T$_{cr}$, as expected for the scaling regime. On turning to the
$^{63}T_1T$ plots, shown in Figs. \ref{QD}a, \ref{QD1}, we see that
the lower crossover temperature, $T_*$, is clearly visible, while,
just as was the case for the 2-1-4 system, the crossover temperature,
$T_{cr}$, at which $^{63}T_1T$ changes slope, is barely discernable.
As we did for the 2-1-4 system, once $T_{cr}$ is identified, we can
determine the doping dependence of both $c'/\alpha$ and $1/\xi$
for these systems, with the results given in Table \ref{dat}

We next consider YBa$_2$Cu$_3$O$_7$, for which measurements of the
Knight shift \cite{Walstedt}, $^{63}T_1$ and $^{63}T_{2G}$ have been carried
out.
The experimental results of Imai and Slichter \cite{Imai:Slichter:7}
for $^{63}T_{2G}$ and  $^{63}T_1T$ in YBa$_2$Cu$_3$O$_7$ bring out two
important points:

\bigskip
\noindent
{\Large$\bullet$} For $T \ge 125 K$  $\omega_{SF}$ and $\xi^{-2}$ display
Curie-Weiss (linear in T) behavior (Figs. \ref{O7:cr}, \ref{O7})

\bigskip
\noindent
{\Large$\bullet$} For $T \ge 125K$ $\omega_{SF} \xi^2 = const$. This behavior,
which has been previously identified as the Gaussian, $z=2$,
scaling regime, could equally well be regarded as a manifestation of
non-universal
mean field behavior \cite{MMP}.

\bigskip
\noindent
What happens at $125K$, where $^{63}T_1T$ stops being linear in $T$? The
results of Walstedt {\em et al}\cite{Walstedt} provide an essential clue.
For their sample, as shown in Fig. \ref{O7:cr},
the bulk spin susceptibility displays only a minor variation with temperature
in YBa$_2$Cu$_3$O$_7$, with a comparatively shallow maximum at 125K.
However, at lower temperatures the variation
of the spin susceptibility becomes
much more pronounced. We therefore identify $T_{cr}$ as $\sim 125K$
for the doping level present in this sample, which may be slightly
on the underdoped side.
Our ansatz that $\xi = 2$ at $125 K$ then permits us to determine
$\alpha$ directly from the $^{63}T_{2G}$ measurements of Imai {\em et al}
\cite{Imai:Slichter:7}; we obtain $\alpha = 15.6 states/eV$, a
value slightly larger than that previously adapted by Thelen and Pines
\cite{Thelen:Pines}, while,
on making use of Eq. \ref{calp}, we find $c' =35 meV$ for this material.

Our conclusion that $T_{cr} = 125 K$ for ``YBa$_2$Cu$_3$O$_7$'' thus
assigns it a place as a full member of the 1:2:3 family, exhibiting
the same magnetic scaling and spin pseudogap behavior as the underdoped
system, Y$_{0.95}$Pr$_{0.05}$Ba$_2$Cu$_3$O$_7$, but with a substantially
lower value of $T_{cr}$. Viewed from this perspective, the $^{63}T_{2G}$
measurements of Imai {\em et al} \cite{Imai:Slichter:7} confirm our
tentative hypothesis that the crossover at $T_{cr}$ is to a non-universal
mean field regime.
In order to examine more closely the doping
dependence of $T_{cr}$ for this family, we need to know the oxygen doping
level of the samples studied by Walstedt {\em et al} \cite{Walstedt}
and Imai {\em et al} \cite{Imai:Slichter:7}. In the absence of detailed
experimental information, we assume these correspond to
YBa$_2$Cu$_3$O$_{6.95}$, although a somewhat lower oxygen doping
level would be equally plausible. We then obtain the results for
the doping dependence of $T_{cr}$ and $(c'/\alpha)$ for the
1:2:3 family shown in Figs \ref{phase} and \ref{ca}.

We next consider YBa$_2$Cu$_4$O$_8$, a material which
possesses the same planar hole concentration as YBa$_2$Cu$_3$O$_{6.8}$,
for which both NQR and NMR measurements of $^{63}T_1$ and
 $^{63}T_{2G}$ have been carried out by Corey {\em et al}
\cite{Corey:Slichter}, with the results given in Fig.
\ref{QD}a,b. They obtain the very important result that
within experimental error the ratio, ($^{63}T_1 T/ T_{2G}$),
is constant between $215 K$ and $450 K$, thus demonstrating
that YBa$_2$Cu$_4$O$_8$ exhibits scaling behavior over a broad
temperature range, between $T_* = 215 K$ and
$T_{cr} \ge 450 K$. Our examination of the Knight shift results
of Zimmerman \cite{Zimmerman} suggests to us that  $\chi(T)$
reaches its maximum value near $500 K$, so that
$T_{\chi} \simeq 500K$; we therefore adopt the value,
$T_{cr} = 470 K$, and use this assignment to obtain
$c'/\alpha = 4.57$, from an extrapolation of the results
for $^{63}T_1$ to this temperature. From the $^{63}T_{2G}$
results of Corey {\em et al} \cite{Corey:Slichter}, we find
$c' =50 meV$, and $\alpha = 10.9 states/eV$ for this material, while
the temperature variation of $\xi$ is given in Table \ref{dat} and
Fig. \ref{cor}.

The final member of the 1:2:3 family for which $^{63}T_{2G}$
measurements have been carried out is YBa$_2$Cu$_3$O$_{6.63}$,
where the measurements of Takigawa\cite{Takigawa} were used
by Sokol and Pines \cite{Sokol:Pines} to establish scaling
behavior, with $T_* \simeq 170 K$. Because high temperature
($T > 500K$) measurements of $^{63}T_1$ $^{63}T_{2G}$ or
$\chi$ are not available for this system, one cannot obtain
$T_{cr}$ from experiment. However, on assuming that $T_{cr}$
continues to vary linearly with oxygen doping, as Fig. \ref{phase}
suggests that it does between YBa$_2$Cu$_4$O$_8$ and
YBa$_2$Cu$_3$O$_{6.95}$, we find $T_{cr} =(840 \pm 150)K$,
with the corresponding results for $c'$, $\alpha$, and
$1/\xi$ given in Table \ref{dat} and Figs \ref{cor}, \ref{alc}.
Our results for the
doping dependence of $c'$ and $\alpha$ show that while
$c'$ decreases with increased oxygen content, $\alpha$
increases in such a way that the product $\alpha c'$ is
nearly constant, a result we examine further in the following
section.

Our determination of the dependence of $T_{cr}$ on oxygen doping
for the 1:2:3 system makes clear the hierarchy of antiferromagnetic
correlations shown in Fig. \ref{cor}, in which ,at a given temperature,
$\xi$ increases rapidly as the planar hole concentration is
reduced. It also enables us to compare 1:2:3 system with the
2:1:4 system. To do so, we choose YBa$_2$Cu$_3$O$_{6.5}$ as
a basis, arguing that up to this oxygen concentration oxygen
atoms added to the insulator, YBa$_2$Cu$_3$O$_6$, go primarily
into chain or interstitial positions, leaving the hole concentration
in the plane near zero. We then assume one hole per plane for
each added oxygen, so that YBa$_2$Cu$_3$O$_{6.95}$ corresponds
to a hole concentration, $n_h = 0.225$, etc, while for the
La$_{2-x}$Sr$_x$CuO$_4$ system we assume that each added $Sr$ atom
introduces a hole in the plane. The resulting dependence of
$T_{cr}$ on hole concentration is given in Fig. \ref{spg}.
We see that within the not inconsiderable uncertainties of
$T_{cr}$ and hole concentration, the two families display the
same dependence of $T_{cr}$ on hole concentration. These
results lead us to two important conclusions:

\bigskip
\noindent
{\Large$\bullet$}
Despite their somewhat different band structure and Fermi surfaces,
the 2:1:4 system and 1:2:3 system possess essentially the same
magnetic phase diagram. Put another way, that diagram is only very
weakly dependent on band structure; it simply reflects the planar
hole concentration.

\bigskip
\noindent
{\Large$\bullet$}
The bilayer couplings\cite{Millis:Monien}
found in neutron scattering experiments on the
1:2:3 system plays little or no role in determining spin pseudogap and
scaling behavior.

\section{Theoretical Model}

Several models have been applied to describe the spin fluctuations
in the metallic state. Two that appear
consistent with experimental picture described above are
the $\sigma$-model\cite{CHN,Chubukov:Sachdev} and
the nearly antiferromagnetic Fermi liquid (NAFL) description
The starting point of the $\sigma$-model is a set of localized
spins interacting with the fermionic background.
It is possible then to integrate out fermions, so that one is
left with the action of the ordering field.
It can be shown that the $\sigma$-model action
Eq.(\ref{smodel}) should then
be generalized to include the effect of
quasiparticle damping, if this kind of damping is allowed by
conservation laws.
This leads to a description of the
$z=1$ to $z=2$ crossover \cite{SCS}.
If the quasiparticle source of damping is absent,
the fermions just change the $\sigma$-model coupling constant
and the spin wave velocity.
This model relies heavily on the existence of the propagating
spin wave excitations in the metallic state, which so far has not
been unambiguously confirmed in the direct experiments. In the NAFL description
such spin wave excitations are not needed, but can arise as a result of a
frequency-dependent effective interaction. The ``spin pseudogap''
is then related to a nonlinear feedback effect which brings about the
equivalent of $z=1$ scaling. We believe
that both models describe from different viewpoints the same physics
of the propagating spin excitations in the metallic underdoped
cuprates. We now consider those in more detail, with particular
attention to the generalization of the MMP model,
Eq(\ref{MMP}), to finite
frequencies, in order to be able to compare the model
prediction with the results of neutron scattering experiments.

\subsection{QNL$\sigma$M plus holes}

In this sub-section we discuss the spin-fermion model. We assume
that the fermionic and spin degrees of freedom can be separated,
and the action for the spins is the $\sigma$-model.
We further assume a specific form for the interaction between the
fermion and spin degrees of freedom:
\begin{equation}
H_{int}=J_0 \sum_{ij} \psi^+_i ({\bf \sigma}_{ij} {\bf S}) \psi_j,
\label{inter}
\end{equation}
where the ${ij}$ are the spin indices.

It can be shown\cite{Liu} that integrating out fermions leads to the following
action in this model \cite{SCS}:
\begin{equation}
S = \frac{1}{2 g}\, T \sum_{\omega_n} \int \frac{d^2 q}{4 \pi^2}
| {\bf n} ({\bf q}, \omega_n)|^2 \left(c_0 q^2 + \frac{\omega_n^2}{c_0}\, +
\gamma
|\omega_n| \right).
\label{action}
\end{equation}
Here both $g$ and $c_0$ are renormalized by fermions. Quasiparticles
also lead to an additional damping, the $\gamma |\omega_n|$ term
in the action, if this kind of damping is allowed by the conservation laws.
The $\sigma$-model constraint $|n|=1$ remains valid
in the this case.

Applicable to systems with two components \cite{Emery},
the decomposition of the one-component system in two components,
the ``spins'' ${\bf S}$
and the ``fermions'' $\psi$ could be artificial. In fact, at high doping
YBa$_2$Cu$_3$O$_{6+x}$  is known to be a one-component
system, where the Fermi surface is measured in
the ARPES experiments \cite{photoemission}. There is no proof that the
model Eq.(\ref{action}) applies in the one-component case, although
some arguments in its favor have been given by Sachdev {\em et al} \cite{SCS}.
The necessary condition for the model Eq.(\ref{action}) to be valid is , in our
opinion, an almost local character of the copper spins.

The scaling analysis of the model\cite{SCS} specified by Eq.(\ref{action})
leads to a magnetic susceptibility near
the ordering wave vector $(\pi, \pi)$
which is a universal function:
\begin{equation}
\chi({\bf q}, \omega) = \frac{Z}{(k_B T)^{- \eta}}\,
\left(\frac{ \hbar c}{k_B T} \right)^2
\Phi_s\left(\frac{q}{T},
\frac{\omega}{T}, \frac{\Delta_0}{T}, \frac{\Gamma}{T}\right).
\end{equation}
Here $q$ marks the deviation from the ordering wave vector $(\pi,\pi)$.
The two energy scales, $\Delta(T=0)$ and $\Gamma$, reflect the energy gap
in the spin wave spectrum and the strength of the fermionic
damping, respectively. The scaling analysis\cite{SCS} shows that
fermions do not necessarily overdamp and destroy the spin waves.
At low damping the crossover is the same as in the pure
$\sigma$-model. The only difference is in the presence
of a finite fermionic damping in the QD regime. At higher $\Gamma$,
there are two crossovers with decreasing temperature. The first one,
from the high-temperature QC $z=1$ to the low-temperature QC $z=2$
regime, happens at $T \sim \Gamma$. At this temperature, according
to this analysis, the temperature dependence of the correlation length
$\xi \propto 1/T$ should change to $\xi^2 \propto 1/T$. The lower second
crossover at $T \sim \Delta^2/\Gamma$  corresponds to the onset of
the QD $z=2$ regime.

As is the case for the pure $\sigma$-model \cite{Chubukov:Sachdev},
the $\sigma$-model with fermionic damping,
Eq.(\ref{action}), can be approached using a $1/N$ expansion technique.
For $N=\infty$, the resulting magnetic susceptibility takes the form
\cite{BPST}:
\begin{equation}
\chi({\bf q}, \omega) =
\frac{\alpha \xi^2}{1 + ({\bf q} - {\bf Q})^2 \xi^2 -
\frac{\omega^2}{\Delta^2}\,
- i \frac{\omega}{\omega_{SF}}}
\label{chi}
\end{equation}
Here $\omega_{SF} \propto \xi^{-2}$ is the characteristic frequency
of the damping in the particle-hole channel, $\Delta = c/\xi$ is
the value of the actual gap in the spectrum of the spin wave excitations
of the disordered state. Eq.(\ref{chi}) can be understood as a
the mean field propagator for the spin excitations damped in
quasiparticle channel\cite{BPST,Gorkov}.
Note, that the only source of damping in mean
field theory is the decay of spin wave into two fermions.
Another source of damping, multiple spin wave scattering,
appears only beyond the mean field, in the first order in $1/N$.
On the contrary, the energy spectrum of the disordered magnons is
already adequate in the mean field approximation\cite{Norbert}.
To summarize these considerations, we therefore write:
\begin{equation}
\frac{1}{\omega_{SF}({\bf q}, \omega)} \, =
\frac{1}{\omega_{Fermion}}\, +
\frac{1}{\omega_{Spin}({\bf q} \xi, \omega/\Delta)},
\label{omsf}
\end{equation}
where $\omega_{Fermion}$ is the fermionic damping frequency appearing in the
mean field theory, while $\omega_{Spin}$ appears as a result of $1/N$
corrections, and displays spatial and frequency dependence on the scale of the
correlation length.

The imaginary part of eq.(\ref{chi}) has the form:
\begin{equation}
\chi''({\bf q}, \omega) = \frac{\alpha \xi^2 \frac{\omega}{\omega_{SF}}}{(1+
q^2\xi^2 -\frac{\omega^2}{\Delta^2})^2 + \frac{\omega^2}{\omega_{SF}^2}},
\label{imch}
\end{equation}
from which we obtain the integrated (or local) magnetic susceptibility:
\begin{equation}
\chi''_L(\omega) =
\pi \alpha \left( \frac{\pi}{2}\, - tan^{-1}\left[
\frac{\omega_{SF}}{\omega}\, - \frac{\omega \omega_{SF}}{\Delta^2}\,
\right] \right).
\label{local}
\end{equation}

Note that for $\omega > \Delta$ the imaginary part of the magnetic
susceptibility Eq.(\ref{imch}) is peaked at an incommensurate
wave vector, which is just a reflection of the spin waves
at high energy. This fact is general and model-independent.
Therefore, it should be seen either in a Fermi-liquid
or a localized spin model.

The spin wave excitations predicted by Eq.(\ref{chi}) need not be well-defined.
A straightforward calculation shows that for ${\bf q} = {\bf Q}$, the
maximum in $\chi''$ occurs at
\begin{equation}
\omega_{max} = \Delta (1-i \gamma),
\end{equation}
where
\begin{equation}
\gamma = \Delta/\omega_{SF},
\end{equation}
from which one obtains:
\begin{equation}
\chi''({\bf q}, \omega_{max}) = \frac{\chi_Q}{\gamma}
\end{equation}
Hence for a well-defined spin wave ($\gamma \ll 1$), one
can find a resonant peak whose magnitude is far in excess of $\chi_Q$,
a situation which is to be contrasted with that which one obtains
when only relaxational behavior is present, and $\chi''_{max} = \chi_Q/2$.
However, as we shall see, in the
normal state of the cuprate superconductors, one likely has
$\gamma \ge 1$. Indeed, if we use the $\sigma$-model result,
$c' =0.55 c$ in the scaling regime, which extends down
to $T_* \lesssim 100 K$ in the systems studied in neutron
scattering experiments, then, at $(\pi, \pi)$,
\begin{equation}
\gamma = \frac{c}{c'} \simeq 1.8, \ \ T_* \lesssim T \lesssim T_{cr}
\end{equation}
We note, that the relation that we have found above, $\alpha c' =const$,
follows directly from the sum rule for the localized spins,
\begin{equation}
\int d^2 q \int d \omega \frac{\chi''({\bf q}, \omega)}{1 - e^{-\omega/T}}\,
\propto \alpha c \propto  \alpha c'
\end{equation}

As applied to the experimental situation described above,
the spin-fermion model provides  a reasonable qualitative description of the
$z=1$ scaling. Based on our
analysis in Section III, we have not found signatures of the
two crossovers in the scaling regime. From this, together with
the results of the measured relaxation rate ratios, which yield $z=1$,
we would conclude that within the $\sigma$-model
 multiple spin wave scattering is
the dominant contribution
to damping of the spin waves in the underdoped materials.
The detailed comparison of this theory with
the inelastic neutron scattering experiments is then obscured by the
fact that the correct form of damping for the spin wave excitations
can only be obtained numerically for the experimentally
accessible temperature and frequency  regions.
The `` spin pseudogap'' temperature in this model is
the temperature where scaling stops working, i.e. the effects of the
natural distance cutoff ($\sim$ the lattice constant, $a$) are appreciable.
It is natural that this happens at certain value of correlation length,
$\xi_{cr} \simeq 2 a$ from Section III. Above the temperature $T_{cr}$
scaling analysis becomes inapplicable. It is impossible to obtain the
precise value of $T_{cr}$, or $\xi_{cr}$ from the scaling analysis alone.
A numerical estimate for $\xi_{cr}$ in the insulator, as mentioned above,
is consistent with ours.

\subsection{Nearly Antiferromagnetic Fermi Liquid Description.}

An alternative approach, the NAFL model\cite{MMP},
started from the high-doping level, where
the magnetic fluctuations have been described as $z=2$ Gaussian
\cite{MMP}. The form eq. (\ref{MMP}) is
obtained by a Taylor expansion of the RPA susceptibility near
the $(\pi, \pi)$ point. The characteristic low-energy mode
in this model is $\omega_{SF} = \Gamma/\xi^2$, with $\Gamma$
of the order of Fermi energy.  The spin waves are
absent, and the spin fluctuations have relaxational character.
This model has been successfully applied to the explanation of
the relaxation rates in YBa$_2$Cu$_3$O$_7$, as discussed in
Section III.

Recently, Monthoux and Pines \cite{Monthoux:Pines} (hereafter MP)
have proposed a phenomenological description of the observed spin
pseudogap and $z=1$ scaling behavior which begins at this ``other
end'', the high-doping limit in which a NAFL approach is the appropriate
starting point.
The NAFL description of magnetic scaling proposed by MP
is based on the twin proposal that the strong Coulomb correlations
between planar quasiparticles give rise to a second energy scale,
$\omega_J \sim$ a few $J$, and that spin pseudogap behavior is a
quasiparticle phenomenon, in which the dynamic irreducible particle-hole
susceptibility, $\tilde{\chi}(q, \omega)$, for wavevectors in the
vicinity of $(\pi, \pi)$ becomes temperature-dependent for
$T \lesssim T_{cr}$. More specifically, MP show that the
propagating term in Eq.(\ref{chi}) can arise from the dependence
on $\omega_J$ of the restoring force, $f^a({\bf q}, \omega, T)$,
which gives rise to NAFL behavior in mean-field theory, in which
case the spin gap is related to $\Delta$ by
\begin{equation}
\Delta = \omega_J (\tilde{\chi}_Q/\chi_Q)^{1/2}=\omega_J \tilde{\xi}/\xi,
\label{mp0}
\end{equation}
where $\tilde{\chi}_Q = \alpha \tilde{\xi}^2$ is the nearly
temperature-independent static irreducible particle-hole
susceptibility.
They also show that
if one writes
\begin{equation}
Lim_{\omega \rightarrow 0} \tilde{\chi}( Q, \omega) =
N_Q(T) \tilde{\chi}_Q \omega,
\label{mp1}
\end{equation}
then in the mean-field approximation,
\begin{equation}
\omega_{SF} = \frac{\tilde{\xi}^2}{N_Q(T) \xi^2},
\label{mp2}
\end{equation}
where $\tilde{\xi} \simeq \pi^{-1}$. Hence if the ``commensurate''
quasiparticle density of states, $N_Q(T)$, obeys the
relations,
\begin{eqnarray}
N_Q(T) & = & \tilde{\chi}_Q \ \ \ \ T > T_{cr} \\ \nonumber
N_Q(T) & = & \frac{\tilde{\xi}^2}{c'}\, \frac{1}{\xi}\, =
\frac{\tilde{\chi}_Q}{\alpha c' \xi} \ \ \ \
T_* \lesssim T \lesssim T_{cr},
\label{mp3}
\end{eqnarray}
then one has the QC $z=1$ scaling behavior,
$\omega_{SF} = c'/\xi$, for $T_* \le T \lesssim T_{cr}$,
while above $T_{cr}$, one recovers non-universal mean field
behavior, with $\omega_{SF} \propto \xi^{-2}$.

As  MP show, since
\begin{equation}
\frac{^{63}T_1 T}{^{63}T_{2G}^2}\, \propto  \chi_Q \omega_{SF}
= \tilde{\chi}_Q/N_Q(T)
\label{mp4}
\end{equation}
to the extent that $\tilde{\chi}_Q(T)$ is weakly temperature dependent,
the temperature dependence of the quantity, $N_Q(T)$, may be directly
obtained from experiment. Moreover, on making our ansatz,
$\xi_{cr} = 2$, then at $T_{cr}$ we have, from Eq.(\ref{mp4}), a
scaling relation between $\alpha$ and $c'$,
\begin{equation}
\alpha c' = \frac{1}{2},
\end{equation}
which is quite close to the values determined in Table \ref{dat}.
In Fig \ref{MP}, we give the results for
$T_{2G}^2/(T_1T)$ for YBa$_2$Cu$_3$O$_{6.63}$, YBa$_2$Cu$_4$O$_8$,
and YBa$_2$Cu$_3$O$_7$  obtained using Eq.(\ref{mp4}) and the experimental
results of Takigawa {\em et al} \cite{Takigawa} and Imai {\em et al}
\cite{Imai:Slichter:8}. It is instructive to
compare the temperature dependence of
$T_{2G}^2/(T_1T) \propto (1/\tilde{\chi}_Q) N_Q(T)$
between $T_*$ and $T_{cr}$ shown in Fig \ref{MP}
for the YBa$_2$Cu$_3$O$_{6.63}$ and YBa$_2$Cu$_4$O$_8$ samples
with that found for $\chi_0(T)$, since at long wavelengths one expects
to find, by analogy with Eq.(\ref{mp1}),
\begin{eqnarray}
Lim_{\omega \rightarrow 0} \tilde{\chi}''(0, \omega) & = &
N_0(T) \tilde{\chi}_0(T_{cr}) \omega \nonumber \\
& \simeq & \chi_0(T) \tilde{\chi_0}(T_{cr}) \omega,
\label{mp5}
\end{eqnarray}
with $N_0(T_{cr}) \simeq \tilde{\chi}_0(T_{cr})$ and
$N_0(T) \propto \chi_0(T)$. We find that
$[N_Q(T)/ \tilde{\chi}_Q(T)]$ and $\chi_0(T)$ display a comparable
temperature variation between $T_*$ and $T_{cr}$; however although
both $[N_Q(T)/ \tilde{\chi}_Q]$ and $\chi_0(T)$ vary linearly with T
over this temperature region, the respective slopes for the
two quantities do not agree.

We further note that the temperature $T_*$ which marks the lower
endpoint of magnetic scaling behavior is barely visible in
Fig. \ref{MP} as the temperature at which
$[N_Q(T)/\tilde{\chi}_Q]$ changes slope. From Eq.(\ref{mp3})
we see that in the NAFL description of the QD state proposed
in MP, at low temperatures, where $\xi \simeq const$, $\omega_{SF}$
varies as $(a + b T)^{-1}$, as is found experimentally,
rather than the exponential behavior
predicted by the $\sigma$-model.

\section{Inelastic neutron scattering results.}

    In this section we discuss the consistency of the
inelastic neutron scattering experiments with the results of NMR
experiments and the scenario developed above.  We start with
YBa$_2$Cu$_3$O$_{6+x}$. Early experiments on this system\cite{Rossat-Mignod}
showed the presence of the gap in the excitations near the antiferromagnetic
wave vector, which increased with increasing doping, consistent with
the scaling model discussed above. However, it has been difficult to
extract the magnetic scattering from the unpolarized neutron experiments
due to the presence of acoustic phonons at similar frequencies.
The quality of the YBa$_2$Cu$_3$O$_{6+x}$
samples and measurements has improved considerably since the early work.
In particular, an analysis of magnetic
excitations\cite{Mook,Fong} has been done
in YBa$_2$Cu$_3$O$_7$ using a polarized neutron scattering technique, which
enables one to extract the phonon part of the scattering cross-section.
Detailed experiments\cite{Sternlieb}
have also been performed
on the underdoped compound YBa$_2$Cu$_3$O$_{6.6}$. In what follows,
we focus on these two  recent experiments.

  Fong {\em at al} \cite{Fong} have recently shown experimentally that,
contrary to the earlier work of Mook {\em et al} \cite{Mook},
the broad peak at $41 meV$ corresponding to a magnetic excitation appears
{\em only} in the superconducting state.
As a result, they argue that this excitation indeed corresponds to
quasiparticle
pair creation in the superconducting state.
Here we wish to suggest an alternative
explanation for the $41 meV$ magnetic peak: {\em that it is a spin wave which
only becomes visible as a distinct excitation in the superconducting state}.
Such an excitation is consistent with the singlet to triplet transition
identified by Fong {\em et al} \cite{Fong}.
As we have remarked in Section IV, the spin wave excitation
is overdamped in the normal state.
It is much better defined in the superconducting state,
because NMR experiments show that $\omega_{SF}$ increases
rapidly for $T < T_c$; indeed, for YBa$_2$Cu$_3$O$_7$,
Martindale {\em et al} \cite{Martindale} find,
$\omega_{SF}(0.75 T_c) \simeq \omega_{SF}(T_c)/10$.
As a result, the gap in the spin excitation spectrum,
$\Delta = c/\xi$, should be observable in inelastic neutron scattering
experiments.
{}From our analysis for YBa$_2$Cu$_3$O$_7$ we obtain $c' \simeq 35 meV$
and $\xi(T_c) \simeq 2.3$.
If $\Delta \sim 41 meV$, then one must have
\begin{equation}
c = \Delta \xi \sim 90 meV
\label{co7}
\end{equation}
As a result, in the normal state, $\gamma = \Delta/\omega_{SF} \simeq 2.5$,
and the spin wave excitations would be unobservable there,
in agreement with the
conclusion of Fong {\em et al} \cite{Fong}. On the other
hand, at $T \simeq 0.75 T_c$, one has $\gamma \simeq 0.25$,
and the excitation becomes readily observable.

We next consider the $\sigma$-model and NAFL interpretation of this result.
In the $\sigma$-model, if the damping of
spin excitations is accounted by multiple spin scattering {\em alone},
the gap in the spectrum of magnetic excitations would be
$\Delta(T_c) \simeq 28 meV$. Since the actual gap is considerably larger,
the fermionic source of damping must be appreciable in YBa$_2$Cu$_3$O$_7$.
Using $\Delta = 1.82 \omega_{Spin} \xi$ and Eq.(\ref{omsf}), we obtain
$\omega_{Spin}/\omega_{Fermion} \simeq 0.47$, as the ratio of the fermionic
and spin sources of damping. Thus, the spin excitation in YBa$_2$Cu$_3$O$_7$
could be a spin wave. As the temperature decreases below $T_c$ the
particle-hole damping disappears.
The gap in the spin wave spectrum causes
the ``spin'' part of damping to also decrease with temperature.
As a result, the spin wave spectrum is much
better defined in the superconducting state.

In the NAFL description,
\begin{equation}
c = \omega_J \tilde{\xi},
\end{equation}
so that $\omega_J \sim 2 J$, a not unreasonable value,
while $\omega_{SF}$ of course reflects only quasiparticle behavior, with its
rapid increase below $T_c$ being {\em entirely} a consequence of the
superconducting gap. Moreover,
measurements of $T_{2G}^{-1}$ in the superconducting
state of YBa$_2$Cu$_3$O$_8$\cite{Corey:Slichter} show that it
gradually decreases below $T_c$, a decrease which, according
to Eq.(\ref{T2x}), directly reflects a decrease in $\xi$.
Since $\Delta = c/\xi$, we would accordingly
expect a corresponding increase in $\Delta$. There are
at present no
measurements of $T_{2G}$ for YBa$_2$Cu$_3$O$_7$; if we
assume that $\xi$ decreases, by some $20 \verb+%+$, from 2.2
at $T_c$ to 1.8 for  $T \lesssim T_c/2$ (by which temperature
the gap is fully established), then a value of $\sim 1.9$ at $0.75 T_c$ is not
unreasonable. On taking $c=74 meV$, we would then find $\Delta = 39 meV$ at
70K,
and $\Delta \simeq 41 meV$ for $T \lesssim 0.5 T_c$, a result in reasonable
agreement with experiment.

     We next estimate the spin wave spectrum in
YBa$_2$Cu$_3$O$_{6.55}$, where detailed
inelastic neutron scattering results  were reported recently
by Sternlieb {\em et al} \cite{Sternlieb}. From our analysis we
find $c' \simeq 70 meV$. In the $\sigma$-model, the ``spin''
source of damping may be expected to be  dominant for
this compound, and therefore $c \simeq 127 meV$.
Since this compound is close to the IM border, and exact doping level
is not known, the correlation length is hard to determine by
extrapolation. To obtain a gap of $\Delta \simeq 10 meV$, which is
experimentally observed, we need $\xi \simeq 12.7$ at low temperature,
a value which would require that the system be at a lower doping
level than YBa$_2$Cu$_3$O$_{6.63}$.
In the NAFL description,
$c$ would not be expected to change appreciably with doping, so that
with $c = 74 meV$, we find a low temperature gap of
$10 meV$ with $\xi =7.4$. This latter length is just what
we find for YBa$_2$Cu$_3$O$_{6.63}$ near
$T_c$, so the NAFL description is clearly
consistent with the emergence of a spin gap at $10 meV$
well below $T_c$.

Sternlieb {\em et al} found that the
local (integrated) spin susceptibility  $\chi''_L(\omega)$
exhibits $\omega/T$ scaling. The temperature at which the scaling breaks
down decreases with increasing $\omega$.
This scaling behavior is generic for the $z=1$ QC regime. The scaling
function for $\chi''_L(\omega/T)$ was found by Chubukov and Sachdev
\cite{Chubukov:Sachdev}. In the region of $\omega$ and $T$ where
the magnetic response was measured, $\chi''(\omega/T) \propto \omega/T$,
consistent with experiment. The decrease of the temperature $T_*$
with increasing energy transfer is also expected on the basis of
scaling arguments.

Provided $\gamma$ is small enough,
the spin wave spectrum that we predict can be observed in neutron
scattering experiments. According to Eq. (\ref{chi}), for
$\omega > \Delta$  $\chi''({\bf q}, \omega)$ is peaked at
the incommensurate wave vector, displaced from $(\pi, \pi)$ by
 $\delta q = \sqrt{\omega^2 - \Delta^2}/c$. However, direct measurement
of the spin wave spectrum in YBa$_2$Cu$_3$O$_{6.6}$ is obscured
by some relatively large degree of imcommensuration observed in
this material\cite{Tranquada}. We estimate the width enhancement due
to the spin waves at $30 meV$ for this material as
$2 \delta q \simeq 0.05 [r.l.u.]$, while the experimental width
of the peak is $\sim 0.25 \pm 0.1 [r.l.u.]$ is independent of energy transfer.
Thus the predicted peak width enhancement agrees within experimental error
with the value reported by Sternlieb {\em et al} \cite{Sternlieb}.

We now consider La$_{1.85}$Sr$_{0.15}$CuO$_4$, a compound where an
extensive neutron scattering study\cite{Mason} of magnetic excitations has been
reported. We note first that the magnetic excitations were
found at incommensurate positions, in contradiction with the one-component
theory of NMR discussed in Section II, which requires that $\chi''({\bf q},
\omega)$
be peaked at a commensurate wave vector. According to Barzykin {\em et
al}\cite{BPT},
the only possible resolution of this contradiction in the framework of the
one-component model lies in the assumption of some sort of domain formation due
to tiny regions of phase separation.
If this is not what happens, more than one copper spin component should be
involved.
We further assume that the one-component theory is correct, and imply the
presence of
some internal superstructure leading to four distinct peaks in the neutron
scattering
experiment. However, we note that the analysis done here is more general than
the
one-component theory reviewed in Section II, since it is based on copper NMR
only,
and the possible incommensurability is not terribly important.
Since $^{63}T_{2G}$ experimental results are not so far available, to our
knowledge,
in the metallic La$_{1.85}$Sr$_{0.15}$CuO$_4$,
it is not possible to determine all the parameters entering
theory. We find that for $\alpha = 15 States/eV$,
$c' \simeq 29 meV$ and $\xi \simeq 7$. Thus, this leads to a predicted spin
gap, $\Delta \simeq 7.5 meV$, for the $\sigma$-model, and $\Delta \simeq 10
meV$ if
$c = 74 meV$ for this compound. We also predict $z=1$ scaling for $T > 90K$.
This temperature $T_*$ should decrease with increasing energy transfer.
None of these effects have been
so far observed in La$_{1.85}$Sr$_{0.15}$CuO$_4$. We found some vague
indication of the spin gap
of $\sim 6 meV$ in the integrated intensity measurements at $35K$, where the
most detailed
experimental study was reported. However, further study at higher temperatures
is
required, in which the kind of scaling plot for the integrated intensity such
as the one used
by Sternlieb {\em et al} in YBa$_2$Cu$_3$O$_{6.6}$ would be extremely useful.
We note that $z=1$ scaling was found by Keimer {\em et al} \cite{Keimer}
in 2:1:4 system at lower doping level,
in the insulating state, as expected from the model discussed above.

The determination made here of the spin fluctuation parameters, $\alpha$
and $\xi$, for the 1-2-3 system enables us to estimate the maximum spin
fluctuation
signal strength $\sim \chi_Q/2$, measurable in inelastic neutron scattering
experiments on the normal state. We find that $\chi_Q$ ranges from $69
states/eV$
for YBa$_2$Cu$_3$O$_7$ to $470 states/eV$ for  YBa$_2$Cu$_3$O$_{6.63}$;
this provides a partial explanation of the difficulty (compared to $O_{6.63}$)
in carrying out quantitative measurements on the fully-doped members of this
family.
As we have noted above, in the supercondicting
state, once quasiparticle excitations can no longer damp the spin gap
excitation at $\Delta$, a considerably shorter signal,
$\sim (\chi_Q/ \gamma)$, where $\gamma \lesssim 1/3$, emerges from the
continuum. We can carry out a complete estimate for $\chi_Q$ for the
2:1:4 system if we assume that for this system as one finds a result
similar to that found for the 1:2:3 system, $c' \alpha \sim 0.54$. We then
find, through interpolation between our results for the $Sr_{0.13}$ and
$Sr_{0.15}$ samples, that $\chi_Q \sim 790 States/eV$ for
La$_{1.86}$Sr$_{0.14}$CuO$_4$ and $\chi_Q \sim 2400 States/eV$ for
La$_{1.9}$Sr$_{0.1}$CuO$_4$. thus one can anticipate significantly
stronger spin fluctuation signals in the 2:1:4 system. It is interesting
to note that the value of the correlation length at $T_*$ for
La$_{1.86}$Sr$_{0.14}$CuO$_4$ is $\xi \sim 8$, comparable to that found
by Mason {\em et al} \cite{Mason} in their neutron scattering experiments on
this
system.

We do not attempt to compare, in this paper, the correlation length
obtained in our analysis of the NMR experiments in YBa$_2$Cu$_3$O$_{6+x}$
system with that measured in the inelastic neutron scattering experiments.
Part of the reason is that the observed peak at the commensurate wave vector
is very broad, a result which would lead to an oxygen NMR spin lattice
relaxation rate which is in disagreement with experiment. On the other hand,
the
neutron scattering experiments of Tranquada {\em et al} \cite{Tranquada} in
YBa$_2$Cu$_3$O$_{6.6}$ have suggested the presence of an unresolved four-peak
superstructure. This superstructure, if present, would make such a comparison
meaningless.

\section{Conclusion.}

   The ``spin pseudogap'' effect in the cuprate superconductors is a
puzzling phenomenon which corresponds to a reduction of the
effective density of states.
We show that this effect is present,
and has similar character, in both single-plane
La$_{2-x}$Sr$_x$CuO$_4$ and in  bilayer YBa$_2$Cu$_3$O$_{6+x}$
families. As a result, we have pursued a single-plane
scenario for this effect in which any
coupling between the bilayers does not play a crucial role.
We find in this scenario that the development of the ``spin pseudogap''
corresponds to the onset of scaling at a certain value of correlation
length. The value of the dynamical critical exponent in the scaling
regime is $z=1$.

We have analysed a large collection of experiments using the
scaling theory (the QNL$\sigma$-model) and the NAFL approach
The scaling theory is based on the existence of
spin wave excitations in the metallic cuprates, which have a finite
energy gap at $(\pi, \pi)$ and a linear dispersion law at
high frequency. It should be
emphasized, that there is no gap other than the gap of the spin wave
excitations in the $\sigma$-model.
This gap should be observable in
inelastic neutron scattering experiments in the superconducting state,
and may have been seen as the 41 $meV$ excitation in YBa$_2$Cu$_3$O$_7$.
The ``spin pseudogap''
observed in the bulk magnetic susceptibility or specific heat measurements is
of different character, it {\em does not necessarily correspond}
to a gap in the quasiparticle spectrum.
We relate the observed spin pseudogap behavior to
the onset of scaling behavior at high temperatures.
It should be noted that a scenario in which the spin pseudogap phenomenon
is related to a failed SDW gap\cite{Kampf:Schrieffer} is also possible.
This mechanism is somewhat similar to the NAFL framework, considered in
Section IV, in which novel physics arises if one takes into account
vertex corrections brought about by stroing magnetic correlations.

In summary, we have applied the $z=1$
Nearly Antiferromagnetic Fermi Liquid Scaling
theory to the analysis of the bulk susceptibility, NMR $^{63}T_1$ and
$^{63}T_{2G}$ relaxation rates, specific heat,
resistivity, Hall effect, and inelastic neutron scattering experiments
in the La$_{2-x}$Sr$_x$CuO$_4$
and YBa$_2$Cu$_3$O$_{6+x}$ families.
The magnetic phase diagram
for the two compounds display the same behavior, and can be
superposed on one another, when plotted as a function of hole
concentration.
Two magnetic crossovers are evident from experimental data. The
upper crossover, at $T_{cr}$,
corresponds to the onset of the $z=1$ scaling behavior. We argue
that above $T_{cr}$ the spin dynamics is non-universal, i.e. influenced
by the lattice cutoff. As a result, no definite predictions can be made
on the basis of scaling theory for $T > T_{cr}$.
Below $T_{cr}$ scaling applies, and
universal scaling functions can be computed.
The temperature $T_{cr}$ decreases with
increasing doping.
The lower crossover temperature , $T_*$, marks the low-temperature
end of the $z=1$ Quantum Critical scaling behavior, and its appearance
is predicted
by the scaling theory.  The magnetic susceptibility, the NMR relaxation rates
$^{63}T_1T$, $^{63}T_{2G}$, and the resistivity are linear in the Quantum
Critical regime between $T_*$ and
$T_{cr}$. Both crossovers disappear in the overdoped case,
as the correlation length becomes too short for $z=1$ scaling to apply.
These considerations are summarized on Figs. \ref{phase}, \ref{phase:La}.
There is no region of applicability for $z=2$ Fermi
Liquid scaling on our magnetic phase diagram in the normal state.
The observed ratio, $^{63}T_1T/^{63}T_{2G}^2=const$ in YBa$_2$Cu$_3$O$_7$,
in our interpretation, corresponds to a mean field non-universal
behavior.

On making the ansatz that $\xi_{cr} = 2$ at the critical
temperature $T_{cr}$ which marks  the onset of the $z=1$ QC regime, we have
determined, from the NMR experiments,
the correlation length as a function of
temperature for both YBa$_2$Cu$_3$O$_{6+x}$ and La$_{2-x}$Sr$_x$CuO$_4$
families, and the spin wave velocity in YBa$_2$Cu$_3$O$_{6+x}$ as a
function of doping.  From this analysis, we can approximately obtain
the energy spectrum of the spin waves, which should be observed in
inelastic neutron scattering experiments on the superconducting
state. Thus, in our scenario,
the peak in magnetic inelastic neutron scattering in YBa$_2$Cu$_3$O$_7$
at $41 meV$, and the spin gap of $10 meV$ observed in YBa$_2$Cu$_3$O$_{6.6}$
could  originate in spin waves. The observation of the full spin wave
spectrum is obscured in these materials by strong phonon scattering and
by some degree of discommensuration found in  YBa$_2$Cu$_3$O$_{6.6}$.

In arriving at our magnetic phase diagram, we have frequently had to
resort to interpolation and extrapolation from existing
experimental data. Our calculations are, in many cases, directly subject
to experimental test. It is our hope and expectation that the NMR experimental
community will now fill in the interstices, and so make more precise
the dependence of $T_{cr}$ and $T_*$ on hole concentration. For example,
detailed measurements of $^{63}T_{cr}$, $^{63}T_1$ and
$\chi_0(T)$ carried out on the same samples, for
samples of known oxygen concentration in the vicinity of
YBa$_2$Cu$_3$O$_{6.95}$, can not only determine
$T_{cr}$ in this region, and establish its dependence on
hole concentration, but can also verify whether, as we have conjectur, the
ratio,
$^{63}T_1 T/ ^{63}T_{2G}$, becomes independent of temperature below
$T_{cr}$. In similar vein more detailed measurements on the Knight
shift and $^{63}T_1$ of the overdoped 2:1:4 system will help one
to deduce $T_*$ and $T_{cr}$ for this system.

We are especially grateful to Takashi Imai, Charlie Slichter, and Bob Corey
for communicating their data prior to publication and for
numerous discussions of the above topics.
We also thank Lev Gor'kov, A. Millis,
Q. Si, A. Sokol, B. Stojkovich and Y. Zha for the
discussions of these and related subjects.
This work was supported by NSF
grant No DMR91-20000 through the Science and Technology Center
for Superconductivity.

%************************************************************

\begin{figure}
\caption{The phase diagram of the $\sigma$-model -- from
Chakravarty {\em et al}\protect{\cite{CHN}}}
\label{phsigma}
\end{figure}

%************************************************************

%***********************************************************************

\begin{figure}
\caption{The magnetic phase diagram determined from NMR experiments
on the YBa$_2$Cu$_3$O$_{6+y}$ system. The temperature $T_{cr}$ marks the
crossover from non-universal behavior to spin pseudogap behavior, while $T_*$
marks the crossover from QC($z=1$) scaling behavior to QD behavior.
Scaling behavior is thus found between $T_*$ and $T_{cr}$.}
\label{phase}
\end{figure}

%***********************************************************************
%***********************************************************************

\begin{figure}
\caption{The magnetic phase diagram determined from NMR and bulk
susceptibility experiments on the La$_{2-x}$Sr$_x$CuO$_4$ system.
The solid circles represent values of $T_{cr}$ determined by us from
NMR experiments; the triangles correspond to the maximum in the bulk
susceptibility
reported by Hwang {\em et al} \protect{\cite{Battlogg}}; the open squares are
the corresponding results obtained by Johnston \protect{\cite{Johnston}}}
\label{phase:La}
\end{figure}

%**********************************************************************
%**********************************************************************

\begin{figure}
\caption{The dependence on hole concentration of the temperature,
$T_{cr}$, for the onset of spin pseudogap behavior determined here
for the 2:1:4 and 1:2:3 systems is compared to the characteristic temperatures
determined by Hwang {\em et al} \protect{\cite{Battlogg}} from transport
and bulk susceptibility measurements and the maximum in bulk susceptibility
found by Johnston \protect{\cite{Johnston}}.}
\label{spg}
\end{figure}

%*********************************************************************
%*********************************************************************

\begin{figure}
\caption{Form factors as a function of momentum, for planar
oxygen and copper sites, in units of the hyperfine coupling constant, $B^2$}
\label{fofa}
\end{figure}

%********************************************************************
%*******************************************************************

\begin{figure}
\caption{$T_{cr}$(denoted by arrows) for La$_{1.8}$Sr$_{0.2}$CuO$_4$ and
La$_{1.76}$Sr$_{0.24}$CuO$_4$, as
determined from the endpoint of linear in $T$ behavior and the maximum of
the$^{63}Cu$ Knight shift measured by Ohsugi {\em et al}
\protect{\cite{Kitaoka}}.}
\label{chi:La}
\end{figure}

%******************************************************************
%**********************************************************************

\begin{figure}
\caption{The spin susceptibility inferred from the Knight shift data
for YBa$_2$Cu$_3$O$_{6.63}$, YBa$_2$Cu$_4$O$_8$,
YBa$_2$Cu$_3$O$_7$, and Y$_{1-x}$Pr$_x$Ba$_2$Cu$_3$O$_7$, with
x=0.05,0.1,0.15.}
\label{knight}
\end{figure}

%**********************************************************************
%************************************************************************

\begin{figure}
\caption{The temperature  $T_*$ (denoted by arrows) for the QC-QD crossover,
as determined from the $^{63}T_1T$ measurements of
Ohsugi {\em et al}  \protect{\cite{Kitaoka}} (a) and
Imai {\em et al} \protect{\cite{Imai:Slichter}} (b)
in La$_{2-x}$Sr$_x$CuO$_4$.}
\label{QD:La}
\end{figure}

%***********************************************************************

%************************************************************************

\begin{figure}
\caption{The temperature dependence of the correlation length in the QC
regime, obtained by us for the La$_{2-x}$Sr$_x$CuO$_4$ system.}
\label{cor:La}
\end{figure}

%***********************************************************************

%***********************************************************************

\begin{figure}
\caption{The variation of $c'/\alpha$ with doping in  La$_{2-x}$Sr$_x$CuO$_4$}
\label{ca:La}
\end{figure}

%***********************************************************************
%*************************************************************************

\begin{figure}
\caption{The determination of $T_*$, the QC-QD crossover,
for the three compounds YBa$_2$Cu$_3$O$_{6.63}$,
YBa$_2$Cu$_4$O$_8$ and YBa$_2$Cu$_3$O$_7$:
(a) From $^{63}T_1$ measurements
(b) From $^{63}T_{2G}$ measurements}
\label{QD}
\end{figure}

%**************************************************************************
%********************************************************************

\begin{figure}
\caption{The determination of $T_*$, the QC-QD crossover,
from $^{63}T_1$ measurements on Y$_{1-x}$Pr$_x$Ba$_2$Cu$_3$O$_7$.}
\label{QD1}
\end{figure}

%********************************************************************
%********************************************************************

\begin{figure}
\caption{The crossover behavior at $125 K$ displayed by the spin contribution
to the magnetic susceptibility \protect{\cite{Zimmerman}}
and $^{63}T_1 T$\protect{\cite{Imai:Slichter:7}} in YBa$_2$Cu$_3$O$_7$.}
\label{O7:cr}
\end{figure}

%********************************************************************

%********************************************************************

\begin{figure}
\caption{ The Curie-Weiss temperature dependence for $\xi^2$
determined from
the $^{63}T_{2G}$ measurements of Imai {\em et al}
\protect{\cite{Imai:Slichter:7}} in YBa$_2$Cu$_3$O$_7$.}
\label{O7}
\end{figure}

%*******************************************************************
%*************************************************************************

\begin{figure}
\caption{$c'/\alpha$, determined from the $^{63}T_1$ measurements,
as a function of doping in YBa$_2$Cu$_3$O$_{7-\delta}$.}
\label{ca}
\end{figure}

%*************************************************************************
%*************************************************************************

\begin{figure}
\caption{The antiferromagnetic correlation length in the
YBa$_2$Cu$_3$O$_{7-\delta}$ system, as
determined from the $^{63}T_{2G}$ measurements. For
the $Pr$-doped compound (the line) the correlation length in the QC
regime was determined from the $^{63}T_1T$ measurements.}
\label{cor}
\end{figure}

%*************************************************************************
%********************************************************************

\begin{figure}
\caption{The parameters $\alpha$ (a) and $c'$ (b), determined for the
three compounds YBa$_2$Cu$_3$O$_{6.63}$, YBa$_2$Cu$_4$O$_8$,
and YBa$_2$Cu$_4$O$_7$, where both $^{63}T_1$ and $^{63}T_{2G}$
relaxation time measurements are available.}
\label{alc}
\end{figure}

%********************************************************************
%************************************************************************

\begin{figure}
\caption{$^{63}(T_{2G}^2/T_1T)$ ($ \propto N_Q(T)/\tilde{\chi}_Q$),
extracted from NMR experiments on YBa$_2$Cu$_3$O$_{6.63}$,
YBa$_2$Cu$_4$O$_8$, and YBa$_2$Cu$_3$O$_7$}
\label{MP}
\end{figure}

%************************************************************************

\newpage
\widetext
\begin{table}
\caption{Spin fluctuation parameters determined from fits to NMR experiments.}
\begin{tabular}{cccccccc}
  & & & $c'/\alpha$ & $\alpha$ & $c'$ &  &   \\
 System & $T_{cr}(K)$ & $T_*(K)$ & ($10^{-3} \ eV^2$) & ($states/eV$)
&($meV$) & $\alpha c'$ & $1/\xi$ ($T_* \lesssim T \lesssim T_{cr}$)  \\
\tableline
\tableline
YBa$_2$Cu$_3$O$_{6.63}$ & 840  & 170 & 7.67 & 8.34  & 64 & 0.53  &
$0.052 + 5.93 \times 10^{-4} T$       \\
YBa$_2$Cu$_4$O$_8$ & 470  & 215  & 4.57 & 10.9  & 50 & 0.545 &
$0.068+9.2 \times 10^{-4} T$      \\
YBa$_2$Cu$_3$O$_7$ & 125 & -- & 2.23 & 15.6 & 35 & 0.545 & --   \\
Y$_{0.95}$Pr$_{0.05}$Ba$_2$Cu$_3$O$_7$ & 200 & 115 &
2.83 & -- & -- & -- & $0.32 + 9.0 \times 10^{-4} T$  \\
Y$_{0.9}$Pr$_{0.1}$Ba$_2$Cu$_3$O$_7$ & 215  & 125 & 3.0
& -- & -- & -- &  $0.31 + 8.8 \times 10^{-4} T$    \\
Y$_{0.85}$Pr$_{0.15}$Ba$_2$Cu$_3$O$_7$ & 260 & 130 & 3.18 & -- & -- & --
&   $0.29 + 8.1 \times 10^{-4} T$   \\
La$_{1.9}$Sr$_{0.1}$CuO$_4$ & 680 & 50 & 2.7 &
-- & -- & -- &  $0.024 + 7.01 \times 10^{-4} T$\\
La$_{1.85}$Sr$_{0.15}$CuO$_4$ & 410 & 70  & 1.91 &
-- & -- & -- &  $0.065 + 1.06 \times 10^{-3} T$\\
La$_{1.8}$Sr$_{0.2}$CuO$_4$ & 120 & 65 & 0.86 & -- & -- & --
&  $0.23 + 2.25 \times 10^{-3} T$ \\
La$_{1.76}$Sr$_{0.24}$CuO$_4$ & 100 & 35 & 0.83& -- & -- & --
&   $0.255 + 2.46 \times 10^{-3} T$
\end{tabular}
\label{dat}
\end{table}

\end{document}